\documentclass[prl,twocolumn,showpacs,superscriptaddress,nofootinbib,english]{revtex4-1}
\usepackage{graphicx,dcolumn,bm,epsfig,graphicx,hyperref,amsmath,amssymb}
\usepackage[usenames]{color}
\usepackage{url}

\DeclareFontFamily{OT1}{pzc}{}
\DeclareFontShape{OT1}{pzc}{m}{it}%
            {<-> s * [1.00] pzcmi7t}{}
\DeclareMathAlphabet{\mathscr}{OT1}{pzc}%
                                 {m}{it}

\hypersetup{
    colorlinks=true,
    linkcolor=red,
    citecolor=blue,
}

\newcommand{\fR}{f_{R}}
\newcommand{\dfR}{{\delta}f_{R}}

\newcommand{\remove}[1]{}

\def\ie{{\frenchspacing\it i.e.}}
\def\eg{{\frenchspacing\it e.g.}}

\def\be{\begin{equation}}
\def\ee{\end{equation}}
\def\ba{\begin{eqnarray}}
\def\ea{\end{eqnarray}}

\begin{document}

\title{Testing Gravity using the Environmental Dependence of
Dark Matter Halos}

\author{Gong-Bo~Zhao}
\affiliation{Institute of Cosmology \& Gravitation, University of
Portsmouth, Dennis Sciama Building, Portsmouth, PO1 3FX, UK}

\author{Baojiu~Li}
\affiliation{DAMTP, Centre for Mathematical Sciences, University
of Cambridge, Wilberforce Road, Cambridge CB3 0WA, UK}
\affiliation{Kavli Institute for Cosmology Cambridge, Madingley
Road, Cambridge CB3 0HA, UK}
\author{Kazuya~Koyama}
\affiliation{Institute of Cosmology \& Gravitation, University of
Portsmouth, Dennis Sciama Building, Portsmouth, PO1 3FX, UK}


\begin{abstract}
In this {\it Letter}, we investigate the environmental dependence of dark
matter halos in theories which attempt to explain the accelerated
expansion of the Universe by modifying general relativity (GR).
Using high-resolution $N$-body simulations in $f(R)$ gravity
models which recover GR in dense environments by virtue of the
chameleon mechanism, we find a significant difference, which depends on the environments, between the lensing and dynamical masses of dark matter halos. This environmental dependence of the halo
properties can be used as a smoking gun to test GR
observationally.
\end{abstract}

\pacs{95.30.Sf, 04.50.Kd, 95.36.+x, 98.80.Jk}


\maketitle

One of the biggest challenges in cosmology is to explain the
recently observed accelerated expansion of the universe. The
acceleration might originate from either ``dark energy" within the
framework of GR, or from a large-scale modification to GR without
introducing new matter species. It could be difficult to
distinguish between these two scenarios by merely measuring the
expansion rate of the Universe, and one has to study the growth of
structure formation in the Universe to break the degeneracy. On
large scales, it is possible to perform model independent tests of
GR by combining various cosmological observations
\cite{Song:2008vm, Zhao:2009fn,Jain:2010ka}, 
but information on linear scales is
limited due to theoretical degeneracies as well as statistical and
systematic uncertainties in observations.

There is ample information available about cluster-scale structure
formation, but it is difficult to predict the observables on
non-linear scales in modified gravity (MG) models. If GR is
modified on large scales, there may appear in gravity new scalar degree of
freedoms, (dubbed {\it scalaron}), modifying GR even
on cluster scales. In order to evade the stringent constraints on
deviations from GR in the solar system, we need a mechanism to
recover GR on small scales by screening this scalar mode. In such
a mechanism, \eg~the chameleon mechanism in $f(R)$ gravity
\cite{Khoury:2003rn, Li:2007}, the mass of this scalar mode
depends on the local density of matter, becoming heavier in denser environments and thereby suppressing 
the scalar interaction. This
environmental dependence can provide us with a smoking gun
for alternative theories to GR \cite{Hui:2009kc, Jain:2011fv}.

In this {\it Letter} we investigate and quantify for the first time the environmental dependence of the difference between lensing and dynamical masses for dark matter halos. Our analysis is based on our high-resolution $N$-body simulations for the $f(R)$ gravity model \cite{Zhao:2010qy}, where the Einstein-Hilbert action in GR is extended to be a general function of the Ricci scalar \cite{fR_review}.

\begin{figure*}[t]
\includegraphics[scale=0.3]{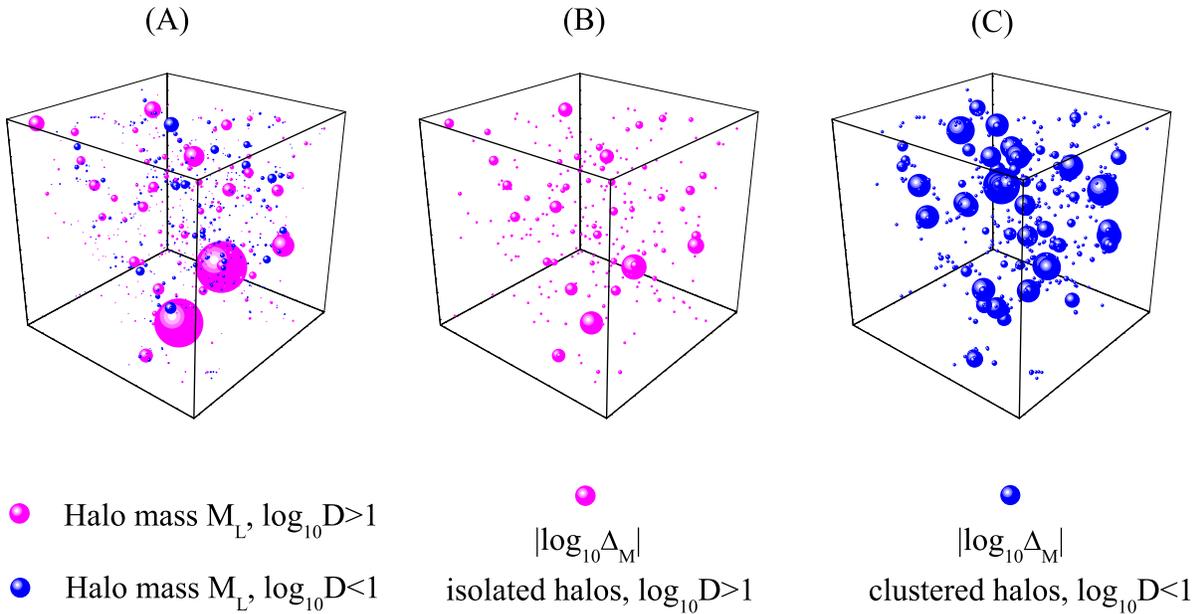}
\caption{The illustration of the three-dimensional distributions of the halos (panel A) and the mass difference $\Delta_M$ for each halo (B,C) for the $|f_{R0}|=10^{-6}$ model. The sizes of the bubbles are proportional to the halo mass $M_L$ in (A), and $|{\rm log}_{10}\Delta_M|$ in (B,C). 
In all the panels, the pink and blue bubbles illustrate the isolated halos, and the halos living in the dense environment respectively. See Eq. (\ref{eq:D}) for the definition of $D$, which quantifies the environment.}\label{fig:3D}
\end{figure*}

In the Newtonian gauge in a general perturbed
Friedmann-Roberston-Walker universe, the line element can be
written as ${\rm d}s^2=a^2(\eta)[(1+2\Phi){\rm
d}\eta^2-(1-2\Psi){\rm d}\vec{x}^2]$ where $\eta$ is the
conformal time, $a(\eta)$ is the scale factor, $\Phi$ and $\Psi$
are the gravitational potential and the spatial curvature
perturbation respectively. The Poisson equation reads \be
\label{eq:poisson} \nabla^2\Phi = 4\pi{G}a^2\delta\rho_{\rm eff},
\ee where $G$ is Newton's constant, and $\delta\rho_{\rm eff}$ is
the perturbed total effective energy density, which contains
contributions from matter and modifications to the Einstein tensor
due to MG. The dynamical mass $M_D(r)$ of a halo is defined as the mass
contained within a radius $r$, inferred from the gravitational
potential felt by a test particle at $r$. It is given by $
M_{D}\equiv\int{a^2}\delta\rho_{\rm eff}dV$, in which the integral
is over the extension of the body. Under the assumption of
spherical symmetry, the Poisson equation can be integrated once to
give \be \label{eq:M_D} M_D(r)\propto{r^2}
d\Phi(r)/dr. \ee  To measure $M_D$ from our N-body simulation we use the force acting on particles to infer the force acting on each halo as a function of the halo radius and $M_D$ can be obtained using Eq. (\ref{eq:M_D}). Observationally, $M_D$ can be estimated from measurements such as velocity dispersions of galaxies. In $f(R)$ gravity $M_D$ includes the contribution from
the scalaron, which mediates the finite-ranged fifth force within
the Compton wavelength. The mass of the scalaron depends on the
local density of matter, resulting in the
environmentally-dependent modifications to $M_D$.

On the other hand, the lensing mass is determined by the lensing
potential $\Phi_+\equiv(\Phi+\Psi)/2$. In $f(R)$ gravity for
example, $\Phi_+$ satisfies $\nabla^2\Phi_+=4\pi
G{a^2}\delta\rho_{\rm M}$, where $\delta \rho_M$ is the matter density
fluctuation if we assume that the background cosmology is close
to that for $\Lambda$CDM. This is the same equation as in GR, since the scalar
mode does not couple to photons and it does not modify light
propagation \cite{Oyaizu:2008sr}. The lensing mass is defined as
$M_{L}\equiv\int{a^2}\delta\rho_{\rm M}dV$, and is the actual measured halo mass in our simulations. Thus we will use $M_L$ to represent the halo mass throughout.
For a spherically symmetric body we
have \be\label{eq:M_L} M_L(r)\propto{r^2}
d\Phi_+(r)/{dr}. \ee

The lensing mass and the dynamical mass are the same in GR, but
they can be significantly different in MG scenarios. To quantify
the difference, we calculate the relative difference $\Delta_M$
between $M_{L}$ and $M_{D}$ for each halo, $ \Delta_M\equiv
M_{D}/M_{L}-1$. Similar quantity, $\mathscr{g}=\Delta_M+1$, was introduced in Ref.  \cite{Schmidt:2010jr}. 
Combining Eqs (\ref{eq:M_D}) and (\ref{eq:M_L}),
we can rewrite $\Delta_M$ as, \be \label{eq:dM}
\Delta_M(r)=\frac{d\Phi(r)/dr}{d\Phi_+(r)/dr}-1, \ee 
In GR $\Delta_M(r)=0$, while in MG models $\Delta_M(r)$ varies depending on the local density.

We have chosen $f(R)$ gravity as a working example to investigate how
$\Delta_M(r)$ correlates with both the halo mass and the environment, and propose a new method
to test GR based on this correlation. For the analysis we shall use the
high-resolution $N$-body simulation catalogue \cite{Zhao:2010qy}
for a $f(R)$ gravity model, $f(R)=\alpha{R}/(\beta{R}+\gamma)$ \cite{Hu:2007nk} 
where $\alpha=-m^2{c_1},\beta={c_2},\gamma=-m^2,m^2=H_0^2\Omega_{\rm M}$
and $c_1,c_2$ are free parameters. The
expansion rate of the universe in this $f(R)$ model is determined
by $c_1/c_2$, and the structure formation depends on $|f_{R0}|$,
which is the value of $|df/dR|$ at $z=0$, and is proportional
to $c_1/c_2^2$. We tune $c_1/c_2$ to obtain the same expansion
history as that in a $\Lambda$CDM model, and choose values for
$|f_{R0}|$ so that those models cannot be ruled out by current
solar system tests. To satisfy these requirements, we set
$c_1/c_2=6\Omega_{\Lambda}/\Omega_{\rm M}$ and simulate three
models with $|f_{R0}|=10^{-4},10^{-5},10^{-6}$. 

\begin{figure}[t]
\includegraphics[scale=0.13]{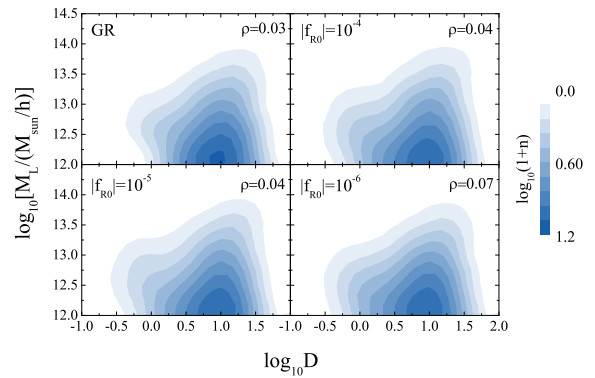}
\caption{The contour plots between halo mass $M_L$ and $D$ for $f(R)$
and GR models on a log-log scale. The shaded colour stands for the
number density rescaled by the average number density of halos in
each pixel on the $M_L$-$D$ plane. }\label{fig:cont}
\end{figure}

\begin{figure*}
\includegraphics[scale=0.22]{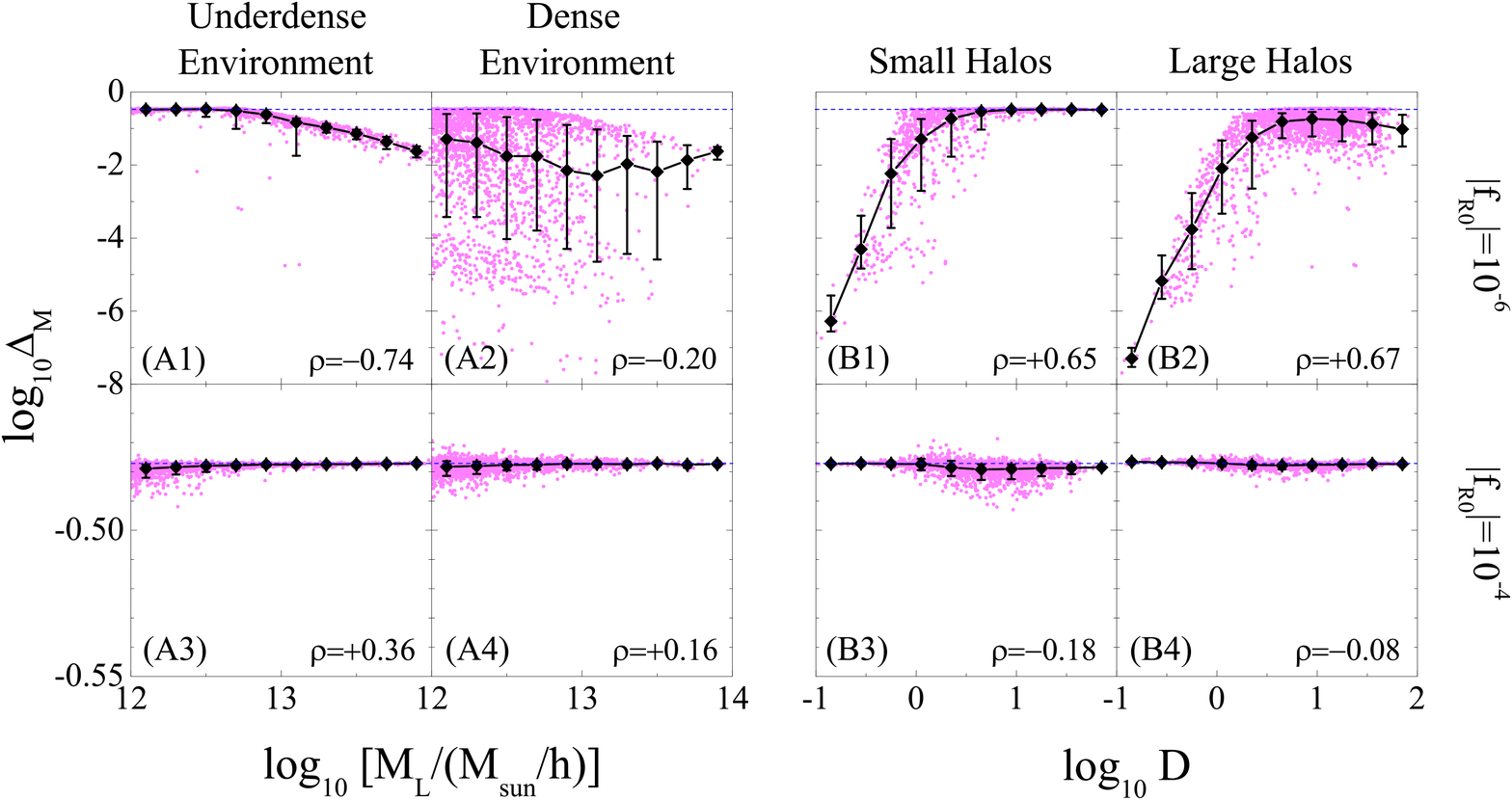}
\caption{$\Delta_M(r_{340})$ as a function of halo mass
(panels $A1$-$A4$) and $D$ ($B1$-$B4$) for our $f(R)$ models on a log-log scale.
Each magenta dot represents one halo in each case, and the black
solid line and the error bars show the mean value of
$\Delta_M(r_{\rm 340})$ and the 1-$\sigma$ error bars
respectively. The correlation coefficient $\rho$ is shown in all cases.}
\label{fig:dM}
\end{figure*}

In $f(R)$ gravity, the effective matter density $\delta\rho_{\rm
eff}$ in Eq (\ref{eq:poisson}) is given by
$\delta\rho_{\rm
eff}=\frac{4}{3}\delta\rho_{\rm
M}+\frac{1}{24\pi{G}}\delta{R}(\fR)$ where $\delta R$ is the
perturbation of the Ricci scalar $\delta{R}(\fR)=-8\pi{G}\delta\rho_{\rm
M}-\frac{3\nabla^2\dfR}{a^2}$ and $\dfR$ is the fluctuation of
$f_R\equiv{df/dR}$. We can see that when the scalar mode vanishes,
\ie~$\dfR=0$, we recover the GR relation between curvature and
matter, $\delta{R}=-8\pi{G}\delta\rho_{\rm M}$ and $\Delta_M=0$. This happens in the dense region where
the chameleon has an effect, but in the
underdense region where $\delta{R}$ can be ignored, $\delta\rho_{\rm
eff}=\frac{4}{3}\delta\rho_{\rm
M}$ so $\Delta_M = 1/3$. One expects a strong correlation between $\Delta_M$ and $M_L$ since halos with large $M_L$ should be `screened' against the modified gravity influence, and GR be locally restored. This has been confirmed by the previous analysis \cite{Schmidt:2010jr,Zhao:2010qy}.

The mass threshold for the screening can be estimated theoretically~\cite{Schmidt:2010jr}. Interestingly, we find that the small halos with masses below the screening mass threshold can also be
well screened if they live in dense environments. This effect is shown visually in Fig.~\ref{fig:3D}. In panel (A) we show the 3-D map of the halo distribution in the $f(R)$ model with
$|f_{R0}|=10^{-6}$ where the size of the bubbles is proportional
to $M_L$, while in panels (B, C) the bubble size is proportional to
$|\log_{10}\Delta_M|$. In other words, a larger bubble means a more massive halo in (A), while it means a better screened halo, in which GR is better restored, in (B, C). In all the panels, the pink and blue bubbles illustrate the isolated halos (log$_{10}D>1$), and the halos living in the dense environment (log$_{10}D<1$) respectively, and these two subsets of halos are complimentary (See Eq. (\ref{eq:D}) for the definition of $D$, which quantifies the environment). Panels A and B look almost the same in pattern, meaning that more massive halos are better screened, and the halo mass is the only factor affecting the screening. This is natural since the environmental effect is removed in panel B by design. On the other hand, in panel C, halos living in dense environments are shown, and the environmental effect is the dominating factor for the screening, so that halos with mass below the screening mass threshold can also be efficiently screened. The difference between the panels (A, C) indicates a clear environmental dependence of $\Delta_M$ -- small halos can be well screened by their neighbouring halos.

This implies that $\Delta_M$ correlates with not only $M_L$, but
also the environment. The environment effect was also noticed by Schmidt \cite{Schmidt:2010jr}. In this {\it Letter}, we shall quantify this effect for the first time. The environmental dependence of $\Delta_M$ provides valuable information
for testing GR, which compliments the information of the mass-dependence of $\Delta_M$. The amount of information can be maximised
if the estimates of the halo mass and environment are
uncorrelated.

The `environment' can be defined such that it suits the physical set-up of the problem, facilitates ease of observations, or both \cite{Haas:2011mt}. For our
purpose, we need an environment indicator which can represent the
local density well, but with least correlation with $M_L$. Such a
quantity was found in Ref.~\cite{Haas:2011mt}, \be\label{eq:D} D_{N,f}\equiv
\frac{d_{N,M_{\rm NB}/M_{L}\geqslant f}}{r_{\rm NB}}, \ee which is
defined for a halo with mass $M_L$ as the distance $d$ to the $N$th
nearest neighbouring halo whose mass is at least $f$ times as
large as that of the halo under consideration, rescaled by the
virialised radius $r_{\rm NB}$ of that neighbouring halo. Clearly, a large value
of $D_{N,f}$ indicates a scarcity of nearby halos, meaning that the
considered halo lives in a low-density environment. It is found
that in GR, $D_{1,1}$ is almost uncorrelated with the halo mass,
and represents the local density well \cite{Haas:2011mt}.

To test the mass-independence of $D_{1,1}$ in the context of
modified gravity, we select the resolved halos from our
high-resolution $f(R)$ and GR simulations with boxsize $B=64$
Mpc$/h$ \cite{Zhao:2010qy}.  In our simulations, the halo mass is measured using $M_L\equiv4 \pi \times N \rho_{\rm crit}  r_{N}^3/3$ where $r_{N}$
is the radius when the density reaches $N$ times of the critical
density of the Universe $\rho_{\rm crit}$, and we choose $N=340$ \cite{Zhao:2010qy}. To be conservative, we only select the well-resolved halos from our simulations, \ie~halos more massive than $10^{12} h^{-1}M_{\odot}$. In
Fig.~\ref{fig:cont}, we show the contour plots between $M_L$ and
$D$, (we will use $D$ to represent $D_{1,1}$ hereafter for brevity), for three $f(R)$ models in comparison with the $\Lambda$CDM model
simulated using the same initial conditions. The darkness of the
shaded colour quantifies the number density of the halos in each
pixel on the $M_L$-$D$ plane. We follow Ref. \cite{Haas:2011mt} to
use the Spearman's rank correlation coefficient $\rho$, which is
the correlation coefficient between the ranked variables and
varies from $-1$ to $1$, to quantify the correlation between $M_L$
and $D$. As we can see, they are almost uncorrelated in all cases
since the absolute value of the correlation coefficient $\rho$ is
much less than unity. This means that the information of the
$\Delta_M$-$D$ relation is highly complimentary to that of the
$\Delta_M$-$M_L$ relation, which provides us with a new means of testing
gravity observationally.

Fig.~\ref{fig:dM} shows $\Delta_M(r_{\rm 340})$ as functions of
$M_L$ and $D$ for two $f(R)$ models. To further
disentangle the residual correlation between $M_L$ and $D$, we
divide the samples into three subsamples in both cases. In the $A$
panels the halos are divided according to
their ordered $D$ values. Halos with $D$ values in the top third of the group (log$_{10}D\gtrsim1$) are classified as halos in an
`Underdense Environment' ($A1,A3$), while those with $D$ values in the lowest third (log$_{10}D\in[-1,0.65]$) are
viewed as halos in an `Overdense Environment' ($A2,A4$). In the $B$
panels, the halos are separated according to their mass, namely,
the halos whose mass is in the top third
($M_L\gtrsim10^{12.7}M_\odot/h$ ) are called `Large Halos'
($B2,B4$), and the third with smallest mass
($M_L\in[10^{12},10^{12.3}]M_\odot/h$) are labeled as `Small
Halos' ($B1,B3$). The horizontal blue dashed line shows
$\Delta_M(r_{340})=1/3$, which is the threshold of $\Delta_M$ in
$f(R)$ gravity.

\begin{figure}[t]
\includegraphics[scale=0.18]{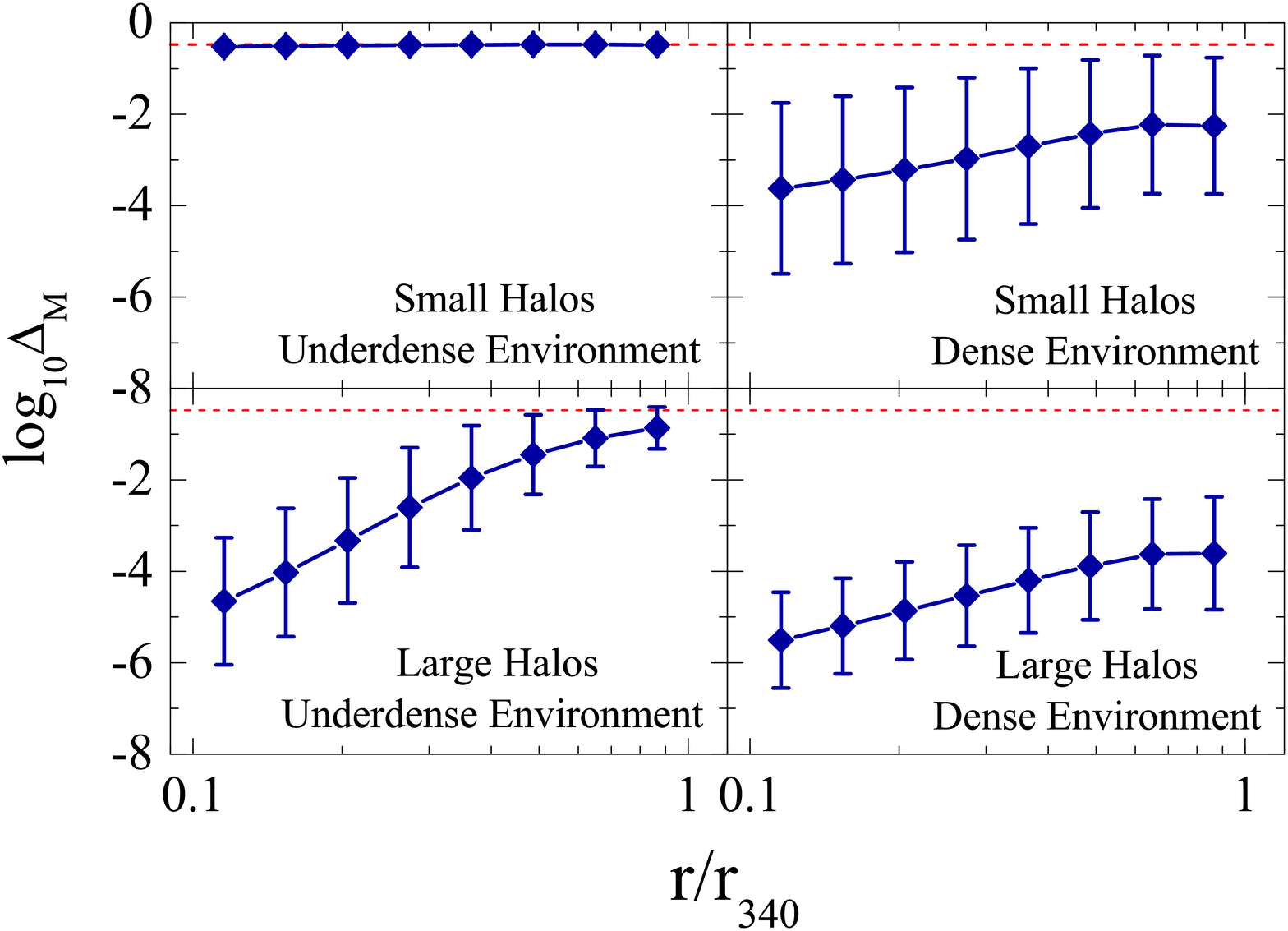}
\caption{The profile of log$_{10}\Delta_M$ as a function of the
rescaled halo radius $r/r_{340}$ for the $|f_{R0}|=10^{-6}$ model.
We show the profile with $1-\sigma$ error bars for the halos
divided into four categories as illustrated in the legend. The red
dashed line shows $\Delta_M(r_{\rm 340})=1/3$.} \label{fig:prof}
\end{figure}

As can be seen from Fig.~\ref{fig:dM}, for the $|f_{R0}|=10^{-6}$
model, $\Delta_M(r_{\rm 340})$ decreases when $M_L$ increases, as
expected. Note that this anti-correlation is stronger
($\rho=-0.74$) in the underdense regions (panel $A1$), as the
environmental effect can be safely ignored in these cases, and
$\Delta_M(r_{\rm 340})$ is mainly determined by $M_L$. In the
overdense environment ($A2$ for the $|f_{R0}|=10^{-6}$ case) for the halos with $D$ values in the lowest third, the effect of external
environment becomes important -- many halos less massive than
$10^{12.3} M_\odot/h$ get screened thanks to the boundary
conditions set by neighbouring halos. An interesting observation
is that some small halos are better screened than the big ones in
this case. This is because many small halos reside in overdense
environments, surrounded by many neighbouring halos or inside
very big halos, while large halos are more likely to be isolated
so that their screening is mainly determined by their mass. For
the $|f_{R0}|=10^{-4}$ model, the halos are very weakly screened
in all cases, and $\Delta_M(r_{\rm 340})$ is close to $1/3$, which
is the maximum relative mass difference in $f(R)$ gravity. The
$|f_{R0}|=10^{-5}$ case is somewhere in between, and is not shown
here.

The environmental effect can be seen more clearly in the
$\Delta_M(r_{\rm 340})$-$D$ plot (panels $B1$-$B4$ in
Fig.~\ref{fig:dM}). For the $|f_{R0}|=10^{-6}$ case, we see a
strong correlation between the two, and this correlation is
largely insensitive to $M_L$ ($\rho\sim0.7$) in both mass bins.
Again, we see that the screening is very efficient in dense
regions even for the least massive halos. The correlation between
$\Delta_M(r_{\rm 340})$ and $D$ for the $|f_{R0}|=10^{-4}$ case is
much weaker, which is because essentially none of the halos are
screened by either their own masses or the environment.

We show the profile of the mass difference as a function of
rescaled radius for the $|f_{R0}|=10^{-6}$ model in Fig
\ref{fig:prof}. To see the environment effect on the profile, we
split the samples according to both the halo mass and $D$ parameter. As we can see, small halos in the underdense region are hardly screened at all, while the halos with similar mass in the dense region
are efficiently screened, and the screening effect is stronger in
the core of the halos. For large halos, the innermost part is
well screened regardless of external environment due to the high
matter density there, but the part close to the edge shows a clear
environmental dependence, and the difference can be as large as 3
orders of magnitude in $\Delta_M$ in different environments. This
is because in this region the external environment plays an
important role.

The lensing mass and the dynamical mass can be measured using strong lensing and the peculiar velocity dispersion measurements,
respectively, and there has been some effort to test GR by
comparing the two observationally
\cite{Bolton:2006yz,Smith:2009fn}. However, the measurements of the
absolute values of $\Delta_M$ are likely to be contaminated by
systematics. Fortunately, the strong environmental dependence of
$\Delta_M$ due to the scalar mode in modified gravity theories may
provide a way to ameliorate this problem.  Observationally, one
could divide the galaxy samples into different groups using $D$,
and measure the difference of $\Delta_M$ among those subsamples.
If a $\Delta_M$-$D$ correlation is found, then it can
be viewed as a smoking gun of a modified gravity signal, which can be independently tested using the $\Delta_M$-$M_L$ correlation.

In this {\it Letter}, we focus on the Chameleon mechanism to recover
GR on small scales. There are different classes of mechanism to achieve
the screening, such as the Vainshtein mechanism \cite{Vainshtein:1972sx} and the symmetron mechanism \cite{Hinterbichler:2010es}.
In the case of the Vainshtein mechanism, it was found that the screening
of halos is almost independent of the environment \cite{Schmidt:2010jr}. Thus the
method we proposed provides not only a new independent test of GR on fully
nonlinear scales but also a way to distinguish between different screening
mechanisms. It is extremely interesting to perform this test using the high-quality
observational data from the upcoming large-scale structure
surveys.

\acknowledgments We thank T. Clemson, R. Crittenden, B. Jain, R. Nichol, L. Pogosian, F. Schmidt and A. Silvestri for discussions. GBZ and KK are supported by STFC grant
ST/H002774/1. BL is supported by Queens' College and DAMTP of
University of Cambridge. KK acknowledges supports from the ERC and
the Leverhulme trust.

\end{document}